\documentclass[aps, prd, twocolumn,.epsf, showpacs, superscriptaddress]{revtex4}

\usepackage{graphicx}
\usepackage{amsmath,amssymb,latexsym}
\usepackage{bm} 

\newcommand{\eg}{\emph{e.g.} }
\newcommand{\ie}{\emph{i.e.} }
\newcommand{\Tr}{\ensuremath{\mathrm{Tr}}}
\newcommand{\Real}{\ensuremath{\mathrm{Re}}}

\makeatletter
\@ifundefined{[}{\newcommand{\[}{\ensuremath{\lbrack}} }{\renewcommand{\[}{\ensuremath{\lbrack}} }
\@ifundefined{]}{\newcommand{\]}{\ensuremath{\rbrack}} }{\renewcommand{\]}{\ensuremath{\rbrack}} }
\@ifundefined{eqref}{\newcommand{\eqref}[1]{(\ref{#1})}}{\renewcommand{\eqref}[1]{(\ref{#1})}}
\makeatother

\begin{document}

\title{The disorder parameter of dual superconductivity in $QCD$ revisited.}

\author{Claudio Bonati}
\affiliation{Dipartimento di Fisica, Universit\`a di Pisa and INFN, Largo Pontecorvo 3, I-56127 Pisa, Italy}
\author{Guido Cossu}
\affiliation{KEK Theory Center, 1-1 Oho, Tsukuba-shi, Ibaraki 305-0801, Japan}
\author{Massimo D'Elia}
\affiliation{Dipartimento di Fisica, Universit\`a di Pisa and INFN, Largo Pontecorvo 3, I-56127 Pisa, Italy}
\author{Adriano Di Giacomo}
\affiliation{Dipartimento di Fisica, Universit\`a di Pisa and INFN, Largo Pontecorvo 3, I-56127 Pisa, Italy}

\begin{abstract}
We discover the origin of the pathologies of the disorder parameter used in previous papers to detect dual superconductivity of 
$QCD$ vacuum, and we remove them by defining an improved disorder parameter.  A check of the approach is made by numerical simulations 
of $SU(2)$ gauge theory, which demonstrate that the approach is consistent and with it that deconfinement is a transition from dual 
superconductor to normal.
\end{abstract}

\maketitle

\section{Introduction} \label{into_sec}

A solid candidate mechanism  for color confinement is dual superconductivity of the vacuum \cite{'tHP, m, 'tH2}. Here 
dual means interchange of electric and magnetic with respect to ordinary superconductors. The idea is that the confining 
vacuum is a condensate of magnetic charges, just as an ordinary superconductor is a condensate of Cooper pairs: the 
chromoelectric field acting between a quark-antiquark pair is thus constrained into Abrikosov flux tubes, whose energy 
is proportional to the length, so that the potential raises linearly at large distances, producing confinement.

The role of magnetic charges in $QCD$ has been actively investigated by the lattice community during the years. The most 
popular approach consisted in detecting monopoles in lattice configurations, and, on the basis of the empirical 
indications that they could be the dominant degrees of freedom \cite{sco,suz}, in trying to extract from the observed 
monopole trajectories an effective lagrangean for them \cite{pol,ts}. Monopole condensation should be readable from 
the parameters of this lagrangean. This program was not convincingly successful. Among others there was the intrinsic 
ambiguity in monopole detection, related to its dependence on the choice of the abelian projection \cite{'tH2}.  
Monopoles are abelian objects, and live in $U(1)$ subgroups of the gauge group: the number and the location of the 
monopoles observed in lattice configurations strongly depend on the choice of the $U(1)$ subgroup. This seems to 
contradict the guess that this choice is physically irrelevant  \cite{'tH2}. This issue was recently clarified 
\cite{bdlp, bdd} by showing that monopoles are gauge invariant objects, as they should for physical reasons, 
but their detection by the recipe of Ref.~\cite{dgt} does indeed depend on the gauge: it was also shown that the so 
called ``maximal abelian projection'' is the one in which the detected monopoles correspond to the real monopoles.  
This legitimates the works performed in the past in the maximal abelian gauge, but maybe the analysis of the effective 
lagrangean could be reconsidered, keeping in mind that monopoles are expected, in principle, to be the dominant 
excitations in the vicinity of the deconfining transition, and not everywhere as assumed in these works.

In Ref.~\cite{bdlp} it was also shown that the Higgs breaking of the magnetic $U(1)$ symmetry coupled to 
monopoles is an abelian projection independent property: if in the confined phase the system is a dual superconductor 
and in the deconfined phase it is normal this is true in all abelian projections.

An alternative approach to the problem has been to define a disorder parameter, to detect magnetic charge condensation 
in the ground state \cite{digz}. The disorder parameter is the vacuum expectation value $\langle \mu\rangle$ of a gauge 
invariant operator $\mu$ carrying non-zero magnetic charge: if $\langle \mu\rangle \neq 0$ vacuum is a dual 
superconductor. In the deconfined phase $\langle \mu\rangle=0$.

A prototype version of the operator $\mu$ is that for the $U(1)$ lattice gauge theory \cite{dp}. The basic idea is to 
define the creation of a monopole as a shift by the classical field of a monopole of the field in the Schroedinger 
representation
\begin{equation}
\mu(\vec x,t) = \exp {\left( i \int d^3y \vec E_{\perp}(\vec y, t)\cdot \vec A^{\,0}_{\perp} (\vec y - \vec x)\right ) } 
\label{mon}
\end{equation}
In whatever gauge we quantize
$\vec E_{\perp}(\vec x, t)$ is the conjugate momentum to the transverse field $\vec A_{\perp}(\vec x, t)$. 
$\vec A^{\,0}_{\perp} (\vec y - \vec x) $ is the field at  $\vec y$ produced by a monopole sitting 
at $\vec x$, in the transverse gauge $\vec \nabla \cdot \vec A^{\,0}_{\perp}=0$. As a consequence
\begin{equation}
\mu(\vec x,t) | \vec A_{\perp}(\vec z, t)\rangle = | \vec A_{\perp}(\vec z, t) + 
\vec A^{\,0}_{\perp} (\vec z - \vec x)\rangle \label{tran}
\end{equation}
Eq.~\eqref{tran} is nothing but the field theory version of the elementary translation
\begin{equation}
\exp {(i pa)} |x \rangle = |x +a \rangle
\end{equation}

A discretized version of the operator $\mu$ can be constructed \cite{dp}, and the order parameter  $\langle\mu\rangle$ 
with it. A theorem can be proved \cite{dp} that such an order parameter is equal to the one defined and discussed 
analytically by completely different methods in Ref.~\cite{fm,cp}. One gets 
\begin{equation}
\langle \mu(\beta) \rangle = \frac {Z(\beta(S+\Delta S))}{Z(\beta S)}\label{zz} 
\end{equation}
with $Z(\beta S)$ the partition function for the gauge field action $S$, $Z(\beta(S+\Delta S) )$ the partition function 
of a modified action, in which the time-space plaquettes at the time $t$, at which the monopole is created, are modified 
to include the electric field term at the exponent of Eq.~\eqref{mon}.
 
It proves convenient for numerical determinations to deal with the quantity
\begin{equation}
\rho(\beta) \equiv  \frac{\partial  \ln  \langle \mu(\beta) \rangle}{\partial \beta}= \langle S \rangle_S - 
\langle S+ \Delta S\rangle_{S+\Delta S} \label{ro}
\end{equation}
from which $\langle \mu(\beta) \rangle $ is easily determined, being $\langle \mu(0) \rangle=1$,
\begin{equation}
\langle \mu(\beta) \rangle = \exp \left(\int^{\beta}_{0}  \rho(\beta')d \beta' \right) \label{exp}
\end{equation}
The numerical determination \cite{dp} of  $ \rho(\beta) $ gave a spectacular demonstration of the theorem proved originally 
in Ref.~\cite{fm} for the Villain form of the action, and extended in Ref.~\cite{cp} to the Wilson form, and at the same 
time a check that the numerical simulation was correct.

\begin{enumerate} 
\item For $\beta < \beta_c$, the critical coupling, $\rho(\beta)$ is volume independent and finite at large volumes, implying, by 
use of Eq.~\eqref{exp}, that $\langle \mu(\beta) \rangle \neq 0$ which means dual superconductivity. 
\item For $\beta > \beta_c$, $\rho(\beta) \approx -|c| L + c'$ grows negative linearly with the size $L$ of the 
lattice, so that, by Eq.~\eqref{exp} $\langle \mu(\beta) \rangle =0$ in the thermodynamic limit.
\item For $\beta \approx \beta_c$ a sharp negative peak is observed, corresponding to a rapid decrease to 
zero of  $ \langle \mu(\beta) \rangle$. The finite size scaling analysis indicates a first order transition.
\end{enumerate}

Notice that $\mu$ is a magnetically charged Dirac-like gauge invariant operator \cite{dp, fm, dirac}, so that 
its vacuum expectation value can be non zero.
 
A similar game was played with the 2d Ising model \cite{cd} where kinks condense in the disordered phase, with the 3d $XY$ model, 
where a phase transition exists related to the condensation of vortices \cite{ddpt}, and with the 3d Heisenberg model, where the 
well known de-magnetising transition can be related to the condensation of a kind of Weiss domains \cite{dmp}.
 
A lattice version of Eq.~\eqref{mon} for non abelian $SU(2)$ gauge theory was then constructed in Ref.~\cite{dlmp} for 
monopoles in specific abelian projections, and in Ref.~\cite{ccos} with a somewhat different formulation. The 
construction was also extended to $SU(3)$ gauge theory \cite{dlmp1}. With the lattice sizes explored 
(up to $4\times 24^3$) and within the statistics reachable at that time, all the features 1), 2), 3) found in the 
prototype $U(1)$ model were reproduced, namely finite, volume independent values for $\rho(\beta)$ at small 
$\beta$'s, a negative linear increase with the spatial size of the lattice at large $\beta$'s and a negative peak at 
the deconfining transition as known from the Polyakov line order parameter, with scaling properties consistent with 
the known order and universality class of the transitions. The disorder parameter proved also to be independent 
of the choice of the abelian projection \cite{3} and to work also in presence of quarks \cite{4}.
Indications for monopole condensation were also obtained by a completely different technique 
in Ref.~\cite{vpc}.
 
An attempt was then made to analyze with the same tool the deconfining transition of the gauge theory with gauge 
group $G_2$ \cite{cddlp}. The question was specially important since the group $G_2$ has no centre and no central 
vortices \cite{pw}, and therefore proving that the observed transition is indeed deconfinement as disappearance of 
dual superconductivity would be an indication that monopoles, as opposed to vortices, be the relevant degrees 
of freedom for color confinement. See also in this connection Ref.'s~\cite{pw1, go}. The formal extension of the 
formalism to deal with monopole condensation in the case of generic gauge groups had been  developed in Ref.~\cite{dlp}.
 
To our surprise it was found \cite{cddlp} that  the expected behavior of $\rho$ was there, but superimposed to a 
negative background  increasing with the spatial size of the lattice at large volumes, in particular at low values of $\beta$, so that 
as a consequence of Eq.~\eqref{exp} the order parameter goes to zero everywhere in the thermodynamic limit, and can 
not distinguish between confined and deconfined phase. We then went back to the old data of Ref.'s~\cite{dlmp, dlmp1} 
and made a more careful analysis with much larger statistics and larger lattices, and also for $SU(2)$ we found a 
similar inconvenience. Similar results were published in Ref.~\cite{gl}, where it was argued about the very possibility 
of defining an order parameter for superconductivity. 
 
In this paper we track the origin of the inconvenience and define an improved order parameter  which is free of 
it, and still is the vacuum expectation of an operator carrying non zero magnetic charge. We check the new order 
parameter numerically for $SU(2)$ gauge theory and find that it satisfies all of the three features listed above.
In order to understand the low $\beta$ background of $\rho$ we study its strong coupling expansion.
 
As for other $d$-dimensional systems showing condensation of $(d-1)$-dimensional topological excitations which had 
been analyzed in the past with analogous techniques we find, by use of strong coupling expansion, that the $U(1)$ gauge 
theory \cite{dp} is exempt of the problem, and so are the $2$-dimensional Ising model, where kinks condense in the high 
temperature phase \cite{cd}, and the $3$-dimensional Heisenberg magnet \cite{dmp}. Instead a non trivial improvement is 
necessary for the $3$-dimensional $XY$ model, where vortices condense in the low temperature phase, and for a generic gauge theory, 
with and without matter fields, and specifically for $G_2$ gauge theory.
 
In this paper we mainly concentrate on the lattice $SU(2)$ pure gauge theory, but the theoretical analysis is valid
for any gauge group. 
 
In Section \ref{impro_sec} we go through  the construction of the disorder parameter, we track the origin of the problem and 
we solve it. To do that we use the strong coupling expansion at low $\beta$. Some details of the expansion are given in 
Appendix \ref{strong_sec}. The origin of the problem is somehow related to the difficulties in defining the disorder parameter 
in presence of electric charges \cite{fmb}.

In Section \ref{mono_sec}  we discuss the choice of the classical field configuration which is added to the quantum fluctuations 
to create a monopole. As a starting point we use the soliton monopole configuration of Ref.'s\cite{'tHm, Pol}.  For the sake 
of clarity algebraic details are presented separately in Appendix \ref{gauge_change_sec}.

In Section \ref{numeric_sec} we present numerical results for $SU(2)$ pure gauge theory on the lattice, demonstrating numerically 
the analysis of Sections \ref{impro_sec} and \ref{mono_sec}.
 
In Section \ref{sub_sec} we present the numerical results of a previous attempt to solve the problem by renormalizing the disorder 
parameter by its value at zero temperature \cite{lat} and show that it is meaningful and consistent with the approach described 
in Section \ref{impro_sec}.
 
Section \ref{concl_sec} contains the conclusions and the outlook.

\section{The improved disorder parameter.}\label{impro_sec}

We start from the $U(1)$ gauge theory for simplicity, recalling the definitions and the notations of Ref.~\cite{dp} to 
make the argumentation self contained.

In the euclidean continuum the disorder parameter Eq.(\ref{mon}) reads
\begin{equation}
\langle \mu \rangle = \frac{\int \prod_{y,\mu} \mathrm{d} A_{\mu}(y) \exp(-\beta (S +\Delta S))}{\int \prod_{y,\mu} 
\mathrm{d}A_{\mu}(y) \exp(-\beta S)}
\end{equation}
where $\beta =\frac{2}{g^2}$, $A_{\mu}$ is the usual vector potential times the electric charge $g$, $S$ is the 
action of free photons
\begin{equation}
S = \frac{1}{4}\int\mathrm{d}^4x \sum_{\mu < \nu} G^2_{\mu\nu}(x)
\end{equation}
and
\begin{equation}
\Delta S = \int \mathrm{d}^3 y \vec E_{\perp}(\vec y, t)\cdot \vec A^{\,0}_{\perp} (\vec y - \vec x)
\end{equation}
On the lattice, assuming the usual Wilson form of the action \cite{w}, the theory is compactified. The partition 
function becomes
\begin{equation}
Z(\beta S) = \int^{2\pi}_0 \prod_{n,\mu}\frac{dA_{\mu}(n)}{2\pi} \exp{(-\beta S)}
\end{equation}
and the action \cite{dp},
\begin{eqnarray}
S &=& \sum_{n, {\mu < \nu}}\Real ( 1 -  \Pi _{\mu \nu}(n)) \label{S_ab} \\
S + \Delta S  & = & \sum_{n, {\mu < \nu}}\Real( 1 -  \Pi'_{\mu \nu}(n)) \label{pp}
\end{eqnarray}
$\Pi _{\mu \nu}(n)$ is the plaquette, the parallel transport around the elementary loop of the lattice in 
the plane $\mu,\nu$, $\Pi _{\mu \nu}(n) = U_{\mu}(n) U_{\nu}(n + \hat \mu) U_{\mu}^{\dagger}(n +\hat \nu)
U^{\dagger}_{\nu}(n)$, $U_{\mu}(n)= \exp (i A_{\mu}(n))$ is the elementary link, $U^{\dagger}_{-\mu}(n + \hat \mu) 
\equiv U_{\mu}(n)$. The lattice spacing has been put conventionally equal to 1. In this notation
\begin{eqnarray}
&& \Pi _{\mu \nu}(n) = \exp(i F _{\mu \nu}(n)) \\
&& \begin{aligned}
F _{\mu \nu}(n)\equiv A_{\mu}(n)+A_{\nu}(n+\hat\mu)\\
-A_{\mu}(n+\hat\nu)-A_{\nu}(n) \label{pi} \label{qqq}
\end{aligned}
\end{eqnarray}
$F _{\mu \nu}(n)$ is the discretized version of the field strength tensor. 

The modified plaquette $\Pi'_{\mu \nu}(n)$, which defines the action $S+\Delta S$ in Eq.~\eqref{pp}, is different from  
$\Pi_{\mu \nu}(n)$ only for electric plaquettes $(\mu,\nu) = (0,i)$ at $n_0= t$, the time at which the monopole is 
created. In formulae
\begin{equation}
\Pi'_{0 i} (t, \vec y)= \Pi_{0 i} (t, \vec y) \exp{(-i A^0_i(\vec y - \vec x))} \label{ppp}
\end{equation}
For all other sites and/or orientations
\begin{equation}
\Pi'_{\mu \nu} (n)= \Pi_{\mu \nu} (n)  \label{pppp}
\end{equation}
One can define $F'_{\mu \nu}(n)$ by the formula $\Pi'_{\mu \nu} (n) =  \exp(i F '_{\mu \nu}(n))$. From 
Eq.'s~\eqref{ppp} and \eqref{qqq} it follows
\begin{equation}
\begin{aligned}
F'_{0 i}(t, \vec n) =  F_{0 i}(t, \vec n) - A^0_i(\vec x - \vec n) = \\ 
=A_0(t, \vec n)+A_i(t,\vec n+\hat\imath)- \\
-A_0(t+1,\vec n)-A_i(t,\vec n)- A^0_i(\vec n - \vec x) \label{fpr}
\end{aligned}
\end{equation}

By the change of variables \cite{dp}
\begin{equation}
A_{i}(t+1,\vec  n) \to A_{i}(t+1,\vec  n) - A^0_i(\vec n - \vec x) \label{pos}
\end{equation}
in the Feynman integral $Z(\beta (S+ \Delta S)) = \int^{2\pi}_0 \prod_{n,\mu}\frac{\mathrm{d}A_{\mu}(n)}{2\pi} 
\exp(-\beta (S + \Delta S))$, the jacobian is $1$ and 
\begin{equation}
\begin{aligned}
F'_{0 i}(t, \vec n) &\to  F_{0 i}(t, \vec n)   \\
F'_{i j}(t+1, \vec n) &\to  F_{i j}(t+1, \vec n) + F^0_{i j} (\vec n -\vec x)\\
F'_{0 i}(t+1, \vec n) &\to  F_{0 i}(t+1, \vec n)  - A^0_i(\vec n - \vec x)
\end{aligned}
\end{equation}
For all the other components and/or sites of the lattice $F'_{\mu \nu}(n_0, \vec n) = F_{\mu \nu}(n_0, \vec n)$ .
 
A monopole has been added at time $t+1$, independent of the spatial boundary conditions (BC) used for the 
quantum fluctuations.
 
By iterating the change of variables at subsequent times the result is that the effect of the operator is to add a 
monopole to the configurations at all times larger than $t$. In particular, if we use periodic spatial BC all of those 
configurations will have the magnetic charge of the added monopole. In terms of states we have
\begin{equation}
Z(\beta (S + \Delta S)) = \langle \mu_{out} | 0 _{in} \rangle \label{muo}
\end{equation}
By $ |\mu \rangle$ we denote the state with one monopole, by $ |\bar \mu \rangle$ that with a antimonopole added to the vacuum state.
 
Starting from Eq.~\eqref{fpr} we could have performed the change of variables
\begin{equation}
A_{i}(t,\vec  n) \to A_{i}(t,\vec  n) + A^0_i(\vec n - \vec x) \label{neg}
\end{equation}
with the result
\begin{equation}
\begin{aligned}
F'_{0 i}(t, \vec n) &\to F_{0 i}(t, \vec n)   \\
F'_{i j}(t-1, \vec n) &\to  F_{i j}(t-1, \vec n) - F^0_{i j} (\vec n -\vec x)\\
F'_{0 i}(t-1, \vec n) &\to F_{0 i}(t-1, \vec n)  -  A^0_i(\vec n - \vec x)
\end{aligned}
\end{equation}
For all the other components and/or sites of the lattice $F'_{\mu \nu}(n_0, \vec n) = F_{\mu \nu}(n_0, \vec n)$.

The change of variables can then be iterated with the result that an antimonopole is present at all negative times, or
\begin{equation}
Z(\beta (S + \Delta S)) = \langle 0_{out} | \bar \mu _{in} \rangle
\end{equation}
 
By a similar technique we can compute the norm of the state $|\mu \rangle $. We can add and subtract to 
$F_{0 i}(t, \vec n)$ the field of the classical monopole. In this case $\Delta S=0$. By operating the change of 
variables Eq.~\eqref{pos} at growing times and the one  Eq.~\eqref{neg} at decreasing times  with the sign of the 
external field changed, we simply get
\begin{equation}
Z(\beta S) = \langle 0_{in} | 0_{out} \rangle = \langle \mu_{in} | \mu_{out} \rangle 
\end{equation}
As a consequence
\begin{equation}
\begin{aligned}
\langle \mu \rangle &\equiv \frac{Z(\beta (S+ \Delta S))}{Z(\beta S)} = \\
&=\frac{\langle \mu_{out}| 0_{in}\rangle}{\sqrt{\langle 0_{out}|0_{in}\rangle}\sqrt{ \langle \mu _{out}| \mu _{in}\rangle}}
\end{aligned}
\end{equation}
Modulo an irrelevant phase  $\langle \mu \rangle $ is the properly normalized probability amplitude from the 
vacuum to the state with a monopole added, and therefore a correct order parameter. The discretization has preserved 
the unitarity of the shift operator Eq.~\eqref{tran}.

At large values of $\beta$, (weak coupling regime) the quantity $\rho$ defined by Eq.~\eqref{ro} can be computed in 
perturbation theory \cite{dp} and is proportional to the lattice size $L$ with a negative coefficient, so that 
$\langle \mu \rangle =0$ in the thermodynamic limit and the vacuum is normal. At low values of $\beta$ we can compute 
it by a strong coupling expansion: if the coefficients of the expansion are finite in the infinite volume limit, 
$\langle \mu \rangle \neq 0$ and the system is a dual superconductor, since the strong coupling series is known to 
have a finite range of convergence in $\beta$.  
The strong coupling expansion of $\rho$ Eq.~\eqref{ro} can be computed with the standard techniques. Many terms cancel 
in the subtraction between the two quantities and the first non-zero term is $O(\beta^5)$ and corresponds to six 
plaquettes containing an elementary cube with an edge parallel to the time axis (see Appendix \ref{strong_sec}). The result is
\begin{equation}
\rho(\beta) = \beta^5 \frac{1}{2^{12}} \sum_{\vec n} \sum_{i<j} [\cos(\alpha_{ij}) -1] +O(\beta^7) \label{sc}
\end{equation}
where $\alpha_{ij} = F^0_{i j}(\vec n)$ is the field strength of the classical field of the monopole, 
$H_k = \frac{1}{2}\epsilon_{kij}F_{ij}$. The sum in Eq.~\eqref{sc} can be approximated by an integral. 
At large distances $\alpha_{ij}$ is small, the cosines can be expanded and 
\begin{equation}
\rho \approx  - \beta^5\frac{1}{2^{13}} \int \mathrm{d}^3x H^2(\vec x) 
\end{equation}
which is convergent since $|H(\vec x)| \propto \frac{1}{r^2}$.  The sum in Eq.(\ref{sc})  is also 
convergent at small distances. The argument is expected to be valid at all orders: indeed at all orders $\rho$ 
is expected to be proportional to $(\cos(\alpha)-1)$  where $\alpha$ is a combination of components of the field 
of the monopole $A^0_{i}(\vec n)$ and since the figures corresponding to the non zero terms of the expansion are closed, 
there will be as many terms with a positive sign as with a negative sign, making the result at large distances go down 
as $\frac{1}{r^2}$ and the result finite. Notice that the result only depends on the magnetic field, which is independent
of the position of the Dirac string.
 
We have presented the case of $U(1)$ in detail to be ready to do a similar analysis for non abelian gauge groups. We 
shall do that for the simplest case $SU(2)$, but everything is general and applies to any group. As a result we shall 
understand and correct the pathologies of the existing disorder parameter quoted in Section I.
 
The definition of the order parameter $\langle \mu \rangle$ of Ref.~\cite{dlmp} for $SU(2)$ is the natural extension to 
the non-abelian case of the construction for $U(1)$ presented above. The original idea is to add the field of an abelian 
monopole to the magnetic field of the residual $U(1)$ of a given abelian projection. Here we shall give a more physical 
view of the construction, showing in particular that the  choice of the abelian projection to start from is completely irrelevant.

Again we have Eq.~\eqref{zz} with
\begin{equation}
Z(\beta S) = \int \prod_{n,\mu}dU_{\mu}(n) \exp{(-\beta S)}
\end{equation}
and Eq.'s~\eqref{S_ab}\eqref{pp} with
\begin{equation}
\begin{aligned}
\Pi_{\mu\nu}(n) \equiv &\frac{1}{\Tr [I]} \Tr \big[U_{\mu}(n)U_{\nu}(n+\hat \mu) \times \\
&\times U^{\dagger}_{\mu}(n + \hat \nu) U^{\dagger}_{\nu}(n)\big]
\end{aligned}
\end{equation}
and $\Pi'_{\mu \nu}(n)= \Pi_{\mu \nu}(n)$ for all $n, \mu ,\nu$ except for the electric plaquettes at $n_0=t$, the time 
at which the monopole is created, which are given by
\begin{equation}\label{pmod}
\begin{aligned}
\Pi'_{i0}(t,\vec n) \equiv \frac{1}{\Tr [I]} \Tr\big[ U_{i}(t,\vec n)U_{0}(t,\vec n+\hat \imath) \times \\
\times M_i(\vec{n}+\hat \imath) U^{\dag}_{i}(t+1,\vec n )U^{\dag}_{0}(t,\vec n)\big]   
\end{aligned}
\end{equation}
$\Tr [I]$ is the trace of the identity, and
\begin{equation}
M_{i}(\vec n ) = \exp\big(i A^0_{i}(\vec n - \vec x)) \label{Mi}  
\end{equation}
$A^0_{i}(\vec x - \vec n)$ is a  classical $SU(2)$ monopole field configuration, \eg the 't Hooft-Polyakov
one, which we will choose with a typical length smaller than the lattice spacing so that its asymptotic form is valid.

We now perform a change of variables in the Feynman integral which defines $Z(\beta(S+ \Delta S))$, 
analogous to that of Eq.~\eqref{pos}, namely
\begin{equation}
U_{i} (t+1,\vec n) \to U_{i} (t+1,\vec n) M_i(\vec n+\hat\imath)\label{cov}
\end{equation}
As a consequence
\begin{align}
\Pi'_{i 0}(t,\vec n) &\to \Pi_{i 0}(t,\vec n)\nonumber\\
\Pi'_{i 0}(t+1,\vec n) &\to \frac{1}{\Tr[I]} \Tr [U_i(t+1,\vec n)U_{0}(t+1,\vec n + \hat \imath)\times\nonumber \\
& \times \bar M_{i}(\vec n+\hat\imath) U^{\dagger}_{i}(t+2, \vec n) U^{\dagger}_{0}(t+1, \vec n)]\label{st}\\
\Pi'_{i j}(t+1,\vec n) &\to \frac{1}{\Tr[I]} \Tr[U_{i}(t+1,\vec n) M_i(\vec{n}+\hat\imath)\times \nonumber \\
& \times U_{j} (t+1, \vec n + \hat \imath) M_j(\vec{n}+\hat\imath+\hat\jmath)\times \nonumber \\ 
& \times M_i^{\dag}(\vec{n}+\hat\imath+\hat\jmath) U^{\dagger}_{i} (t+1, \vec n + \hat \jmath)\times \nonumber \\
& \times M_j^{\dag}(\vec{n}+\hat\jmath) U^{\dag}_{j}(t+1,\vec n)] \label{ss}
\end{align}
In Eq.~\eqref{st} 
\begin{equation}
\bar M_{i}(\vec n) = U^{\dagger}_{0}(t+1, \vec n)M_{i}(\vec n) U_{0}(t+1, \vec n)\label{mbar}
\end{equation}
By repeated use of the Campbell-Baker-Haussdorf formula it is easy to show that, up to terms $O(a^2)$
\begin{equation}
\Pi'_{i j}(t+1,\vec n) \approx \frac{1}{\Tr[I]} \Tr [\exp {i (G_{i j} + G^{0}_{i j})}]
\end{equation}
A monopole has been added at time $t+1$. If we take the field $A^{0}_{i}$ in the unitary representation, which coincides with the 
maximal abelian gauge \cite{bdlp}, and periodic boundary conditions for the quantum fluctuations, the magnetic monopole field at 
large distance will be that of the monopole added, whatever is the abelian projection we started from. The magnetic field 
$F^{0}_{i j}$ is directed along the color direction of the residual $U(1)$ and adds to the corresponding component of the quantum 
field. That component is in any case undefined by terms $O(a^2)$ depending on the conventions used to extract it from the 
configurations, again by the Campbell-Baker-Haussdorf formula.
 
For the same reason the matrix $\bar M_{i}(\vec n)$ of Eq.~\eqref{mbar} has the same $U(1)$ component as $M_{i}(\vec n)$ up to 
terms $O(a^2)$ since the terms $O(a)$ cancel between $U^{\dagger}_{0}(t+1, \vec n)$ and $U_{0}(t+1, \vec n)$. Therefore the change 
of variables Eq.~\eqref{cov} can be repeated at time $t+1$ with the effect of adding a monopole at time $t+2$ and so on. The monopole 
propagates from time $t$ on. The same conclusion follows from the fact that the abelian projected field obeys abelian Bianchi 
identities \cite{bdlp}, so that the magnetic current is conserved \cite{dlp}. Eq.~\eqref{muo} is still valid. There is, however, 
a misuse of language: in the abelian case the monopole propagates unchanged from time $t$ to $\infty$, and is the ``out'' state. 
As we have just seen, in the non-abelian case the monopole configuration gets modified in propagating, even if it preserves 
its quantum numbers, so the configuration is not exactly the ``out'' state.
 
A few remarks:

\begin{enumerate} 
\item All of the equations for the non abelian case also hold for the abelian one where they coincide with those described above 
as can easily be checked.
 
\item In what follows we shall use as classical monopole field $A^{0}_{i}(\vec n)$ the 't Hooft -Polyakov soliton solution in the 
unitary representation: as any magnetic monopole field they decrease as $\frac{1}{r}$ at large distances. The details will be 
presented in the next section.
 
\item As directly visible from the definition Eq.(\ref{pmod}) of $\Pi'_{\mu \nu}$, $Z(\beta (S + \Delta S))$ and $\langle \mu \rangle$ 
are invariant under covariant transformations of  $A^{0}_{i}(\vec n)$,  $A^{0}_{i}(\vec n) \to V(n) A^{0}_{i}(\vec n) V(n)^{\dagger}$ 
but not under gauge transformations of the classical field. We then expect different disorder parameters in different gauges, which 
are however expected to be on the same footing since each of them is the transition amplitude from the ground state to a state with a 
monopole added.  
\end{enumerate}

The disorder parameter of Ref.~\cite{dlmp} was
\begin{equation}
\langle \mu \rangle =\frac{ \langle \mu_{out} | 0_{in}\rangle}{\langle 0_{out}|0_{in}\rangle}  \label{mm}
\end{equation}
The correct quantum-mechanical definition of the disorder parameter, \ie of the probability to find in the vacuum 
state a state with magnetic charge increased by a monopole is
\begin{equation}
\langle \bar \mu \rangle =\frac{\langle \mu_{out} | 0_{in}\rangle}{\sqrt{\langle 0_{out}|0_{in}\rangle} 
\sqrt{\langle \mu_{out} | \mu_{in}\rangle} } \label{mus}
\end{equation}
In the case of the $U(1)$ gauge group we showed that $ \langle \mu_{out} | \mu_{in}\rangle = \langle 0_{out}|0_{in}\rangle$ 
so that Eq.s~\eqref{mus} and \eqref{mm} are the same thing and $\langle \mu \rangle$ as originally defined is the order 
parameter. For non abelian gauge groups we will show that this is not the case. $\langle \mu_{out} | \mu_{in}\rangle$ 
can be written as
\begin{equation}
\langle \mu_{out} | \mu_{in}\rangle = Z(\beta (S + \overline{\Delta S}))
\end{equation}
so that
\begin{equation}
\begin{aligned}
\bar \rho \equiv \frac{\partial \ln \langle \bar \mu \rangle}{\partial \beta} = 
\frac{1}{2} \langle S \rangle_S +  \frac{1}{2} \langle S + \overline{\Delta S} \rangle_{S + \overline{\Delta S}} \\ 
-\langle S +\Delta S\rangle_{S + \Delta S}        \label{ros}
\end{aligned}
\end{equation}
which again coincides with Eq.~\eqref{ro} in the case of $U(1)$, since there $\overline{\Delta S}=0$.
   
We shall now compute $\overline{\Delta S}$ for the generic case and show that it is non trivial: the discrete 
implementation of the operator Eq.~\eqref{mon} proposed in Ref.~\cite{dlmp} does not preserve unitarity. Not only that, 
but both $\langle \mu_{out} | 0_{in}\rangle$ and $\langle \mu_{out} | \mu_{in}\rangle^{\frac{1}{2}}$ tend to zero in the large 
volume limit, in such a way that their ratio stays finite and different from zero at small values of $\beta$.  
   
$\langle \bar \mu \rangle$ of Eq.~\eqref{mus} is the correct order parameter, which will be computed via $ \bar\rho$ of 
Eq.~\eqref{ros} by use of Eq.~\eqref{exp}.
   
We first give the expression for $S+\overline{\Delta S}$. It naturally derives from the definition of scalar product
in the language of path integral.
\begin{equation}
S+\overline{\Delta S} =  \sum _{n,\mu \nu}\Real  [1- \overline{\Pi}_{\mu \nu}(n)]
\end{equation}
where $\overline{\Pi}_{\mu \nu}(n) = \Pi_{\mu \nu}(n)$ everywhere except for the mixed space-time plaquettes 
$\overline{\Pi}_{0i} (t, \vec n)$ at the time $t$ at which the monopole is created, where (see  
Eq.'s~\eqref{pmod}-\eqref{Mi})
\begin{equation}
\begin{aligned}
\overline{\Pi}_{0i} (t, \vec n) = \frac{1}{\Tr [I] } \Tr\big[ U_i(t, \vec n) M^{\dagger}_{i}(\vec n+\hat \imath) \times\\ 
\times U_{0}(t,\vec n + \hat \imath) M_{i} (\vec n+\hat \imath)U_{i}^{\dag}(t+1, \vec n)U_{0}^{\dag}(t,\vec n)\big] \label{sop}
\end{aligned}
\end{equation}
Indeed by the usual change of variables in the path integral which defines $Z(\beta (S+\overline{\Delta S}))$ it can be 
shown that with this definition there is a monopole propagating in the future, and one coming from the past. 
$M_{i}^{\dagger}$ and  $M_{i}$ in the equation \eqref{sop} cancel in the case of $U(1)$ because they commute with the 
links, everything being abelian, and therefore $\overline{\Pi}_{\mu \nu}(n)=\Pi_{\mu \nu}(n)$ everywhere.
   
We shall now compute both $\rho$ and $\bar \rho$ by the strong coupling expansion at low $\beta$'s. The details of 
the expansion are sketched Appendix \ref{strong_sec} below. 
   
Like for the case of $U(1)$ the first non trivial contribution to the strong coupling expansion of $\rho$
is $O(\beta^5)$ corresponding graphically to an elementary cube, and is given by
\begin{equation}
\rho \propto \beta^5 (\Tr [I])^5\sum_{\vec n}(T_{xy} + T_{yz} + T_{zx}  - 3 (\Tr [I])^3) \label{osc}
\end{equation}
and
\begin{equation}
\begin{aligned}
T_{ij}(\vec n) = \Real \Big( \Tr\big[M_{i}(\vec n + \hat \imath)\big] \Tr \big[M_{j}^{\dag}(\vec n + \hat \jmath)\big] \times \\ 
\times \Tr\big[M^{\dag}_{i}(\vec n + \hat \imath +\hat \jmath) M_{j}(\vec n + \hat \imath + \hat \jmath)\big] \Big)
\end{aligned}
\end{equation}
By use of the definition Eq.~\eqref{Mi} we get
\begin{equation}
\begin{aligned}
&\big(T_{ij} -(\Tr [I])^3\big) = (\Tr [I])^3 \Big\[ \cos\big(A^0_{i}(\vec n + \hat \imath)\big) \times \\
&\hspace{2cm}\times \cos\big(A^0_{j}(\vec n + \hat \jmath)\big)\times \\
&\times \cos\big(A^0_{j}(\vec n + \hat \imath +\hat \jmath) -A^0_{i}(\vec n +\hat \imath + \hat \jmath)\big) - 1\Big]
\end{aligned}
\end{equation}
Since, as we shall see in Section \ref{mono_sec}, in the Wu-Yang parametrization of the monopole field 
$|\vec A^{\,0}(\vec x)| \approx_{r\to \infty} \frac{1}{r}$, it is easily seen, by expanding the cosines at large distances, 
that, neglecting finite terms,
\begin{equation} \label{rho_div}
\rho \propto -\beta^5 (\Tr [I])^8\sum_{\vec n} |\vec A^{\,0}|^2(\vec n) 
\end{equation}   
which is linearly divergent to $-\infty$ with the spatial size of the lattice. This is indeed the pathology observed in 
lattice simulations, and what makes $\langle \mu \rangle =0$ at all $\beta$'s in the thermodynamic limit.
   
If we now play the same game with the new order parameter $\bar \rho$ of Eq.~\eqref{ros} we get instead
\begin{equation}
\bar \rho = \rho -\frac{1}{2} \rho_2 \label{ccc}
\end{equation}
with
\begin{equation} 
\rho_2 \equiv  \langle S\rangle_{S} - \langle S + \overline{\Delta S}\rangle_{S+\overline{\Delta S}} \label{ro2}
\end{equation}
In the strong coupling again the first non trivial term corresponds to the elementary cube and gives
\begin{equation} \label{osc2}
\rho_2 \propto -\beta^5 (\Tr [I])^2 \sum _{\vec n}(\bar T_{xy}+ \bar T_{yz}+\bar T_{zx} - 3 (\Tr1)^6 )
\end{equation}
where the coefficient of proportionality is the same as in Eq.~\eqref{osc} and
\begin{equation}
\begin{aligned}
\bar{T}_{ij}(\vec n) = \Big|\Tr\big[M_{i}(\vec n + \hat \imath)\big]\Big|^2 \Big|
\Tr \big[M_{j}^{\dag}(\vec n + \hat \jmath)\big]\Big|^2 \times \\ 
\times \Big| \Tr\big[M^{\dag}_{i}(\vec n + \hat \imath +\hat \jmath) M_{j}(\vec n + \hat \imath + \hat \jmath)\big]\Big|^2
\end{aligned}
\end{equation}
\begin{equation}
\begin{split}
&\bar T_{ij} - (\Tr[I])^6 =(\Tr[I])^6  \times \\
&\hspace{0.5cm}\times \Big\[ \cos ^2\big(A^0_{i}(\vec n + \hat \imath)\big) \cos^2 \big(A^0_{j}(\vec n + \hat \jmath)\big) \times \\ 
&\hspace{0.5cm} \times\cos^2 \big(A^0_{j}(\vec n + \hat \imath +\hat \jmath) -A^0_{i}(\vec n +\hat \imath + \hat \jmath)\big) - 1\Big]
\end{split}
\end{equation}

In computing $\bar\rho$ by use of Eq.~\eqref{ccc} the leading terms at large distances cancel and the final result 
is finite in the thermodynamic limit. 

In formulae
\begin{equation}
\begin{aligned}
&\bar \rho \propto  -\beta^5 (\Tr [I])^8 \sum_{0<i<j}  \Big\[ \cos\big(A^0_{i}(\vec n + \hat \imath)\big) \cos\big(A^0_{j}(\vec n + \hat \jmath)\big) \\ 
&\cos\big(A^0_{j}(\vec n + \hat \imath +\hat \jmath) -A^0_{i}(\vec n +\hat \imath + \hat \jmath)\big) - 1\Big]^2
\end{aligned}
\end{equation}
 
The state $|\mu \rangle $ has zero norm in the infinite volume limit, but its direction in Hilbert space is well defined, 
at least in the confined phase.

This was the leading term in the strong-coupling expansion. The exact result is obtained either by studying systematically 
higher order terms in the strong coupling expansion, or numerically by lattice simulations. This we will do in the next section.
Notice that the coefficient of the first term in the strong coupling expansion depends on the vector potential and not on the 
magnetic field as was the case for $U(1)$ gauge group. This fact is related to the remark  3) above, that the definition of 
$\langle \bar \mu \rangle$ is not invariant under gauge transformations of the classical field of the monopole.

\section{The monopole configuration in non abelian gauge theories}\label{mono_sec}

In the $U(1)$ gauge theory on the lattice the classical monopole configuration by which the field is shifted by the 
monopole creation operator has a Dirac string singularity: $\langle \mu \rangle$ is  fully gauge invariant both with respect to 
regular gauge transformations and with respect to singular gauge transformations like those which displace the Dirac
string \cite{dp, fm}.  There are, however, problems to satisfy in a consistent way the Dirac relation between 
electric and magnetic charge  when electric charges are present \cite{fmb}: this is the case for example in non 
abelian gauge theories where electrically charged fields are necessarily  present, at least the charged partners of 
the $U(1)$ gluon. 

In the non abelian case we shall view the monopole as an $SU(2)$ configuration.

Suppose we want to create a 't Hooft-Polyakov monopole in the unitary representation from the configuration 
with no gluons. 

For our purposes we may consider only the large distance form of the configuration. The notation is standard (see \eg \cite{shnir}).

The monopole configuration in the Wu-Yang form with discontinuity say at a polar angle 
$\theta = \theta_{WY}$ is obtained from the hedgehog gauge configuration (in polar coordinates) 
\begin{eqnarray}
A_{r}=0 \label{ar}\\
A_{\theta} =\frac{\sigma _{\phi}}{2gr}\label{ateta}  \\
A_{\phi} = - \frac{\sigma_{\theta}}{2gr} \label{afi}
\end{eqnarray}
by operating the gauge transformation
\begin{equation} 
U(\theta,\phi) = \exp(-i\frac{\sigma_{\phi}}{2} {\bar \Theta}) {\Lambda} \label{u}
\end{equation}
with
\begin{equation}
\bar \Theta = \theta - \pi \vartheta (\theta - \theta _{WY})\label{tetabar}
\end{equation}
and
\begin{equation}
\Lambda =\exp[-i\frac{ \sigma_{\perp}}{2}( \bar \Theta -\theta)] \label{lambda}
\end{equation}
$\vartheta$ is the usual Heaviside function. The requirement in the definition of $\Lambda$ is that $\sigma_{\perp}$ is $\phi$ and 
$\theta$ independent and that $\Lambda^{\dagger} \sigma_3 \Lambda = \exp i(\bar \Theta - \theta)\sigma_3$. 
$\sigma_{\perp}$ can be \eg $\sigma_1$, or $\sigma_2$ or any fixed combination thereof. The field in the new gauge 
is computed  by use of the formula 
\begin{equation} 
\vec {\tilde A} = U(\theta,\phi)^{\dagger} \vec A U(\theta,\phi) -\frac{i}{g} U(\theta,\phi)^{\dagger}\vec \nabla 
U(\theta,\phi) \label{A}
\end{equation}
The details of the computation are given in  Appendix \ref{gauge_change_sec}. The result is
\begin{eqnarray}
{\tilde A}_r &=&0\label{Ar} \\
{\tilde A}_{\theta}&=&\frac{\pi \delta(\theta -\theta_{WY})}{2gr} \Lambda ^{\dagger} ( \sigma_{\phi} + \sigma_{\perp}) \Lambda \label{Ateta}\\
{\tilde A}_{\phi}&=&\frac{(1- \cos{\bar \Theta})}{2gr\sin\theta} \Lambda ^{\dagger} \sigma_3 \Lambda \label{Afi}
\end{eqnarray}
The non abelian magnetic field is radial 
\begin{equation}
 H_r = \frac{1}{r \sin\theta}[\partial_{\theta}( \sin\theta  {\tilde A}_{\phi}) - \partial_{\phi}  {\tilde A}_{\theta}] +ig [{\tilde A}_{\theta},{\tilde A}_{\phi}] \label{mf}
\end{equation}
As shown in  Appendix \ref{gauge_change_sec} the singularities cancel and the field is simply the coulombic 
field $\vec H = \frac{\hat r}{gr^2} \frac{\sigma_3}{2}$. Moreover the component ${\tilde A}_{\phi}$ of the field in the Wu-Yang 
unitary gauge is continuous inside $SU(2)$ as shown in Eq.~\eqref{ub}. It is instead discontinuous if viewed as an $U(1)$ configuration.
In the limit $\theta_{WY} = \pi$ the singularity becomes the usual Dirac string \cite{shnir}. To create a monopole we should in 
principle create, together with the abelian diagonal part of the field, a singular charged gluon field $\tilde A_{\theta}$ 
which cancels the singularities. For the matrix element of the monopole operator which we are considering (creation from a zero 
gluon configuration) this procedure amounts to keep only the diagonal part  and neglect its 
singularities (Dirac string or  Wu-Yang transfer matrix).  This is the recipe that will be implemented on the lattice 
(Section \ref{numeric_sec}).

If we act with the monopole operator on a generic configuration with spatial periodic boundary conditions, \emph{i.e.} with net 
magnetic charge zero, the magnetic field at large distances will be the one that we add, and we are therefore automatically  
in the maximal abelian or unitary gauge \cite{bdlp}. The singularities cancel anyhow in the magnetic field, except 
possibly for  the  bilinear term in Eq.~\eqref{mf} with the singular classical field ${\tilde A}_{\theta}$ 
Eq.(\ref{Ateta}) and the quantum fluctuation of $\tilde A_{\phi}$, which should however average to zero. Again the recipe is 
to keep only ${\tilde A}_{\phi}$ at $\theta \neq \theta_{WY}$.
 
How this recipe extends to more complicated matrix elements or correlation functions is an open  question which 
deserves further study. The emerging indication is that, contrary to a generic theory containing charged fields, 
in a gauge theory the problem raised in Ref.~\cite{fmb} is structurally solved. Monopole configurations are non 
singular: what produces the singularity is the abelian projection. In the language of Ref.~\cite{bdlp} monopoles in a 
non-abelian gauge theory are violations on non-abelian Bianchi identities, which produce violations of abelian Bianchi 
identities only after abelian projection (Sect. 2 of Ref.~\cite{bdlp}).

\section{Numerical results}\label{numeric_sec}

In this section we present the numerical results obtained by simulating the $SU(2)$ lattice gauge theory. Since both the 
actions $S+\Delta S$ and $S+\overline{\Delta S}$ are linear in each of the link, the usual combination of heatbath 
\cite{creutz80,kp} and overrelaxation \cite{creutz87} can be used for the update of the configuration, even when the 
monopole operator is present. Numerical simulations have been performed on GRID resources provided by INFN.

Our aim is to show that the behaviour of the improved operator $\bar\mu$ in the $SU(2)$ case is similar to the one 
observed in the $U(1)$ lattice gauge theory, namely we want to verify that $\bar\rho$
\begin{itemize}
\item is finite in the thermodynamical limit in the low temperature phase ($\beta<\beta_c$);
\item develops a negative peak in correspondence of the deconfinement transition which scales with the appropriate 
critical indices;
\item diverges to $-\infty$ with the spatial lattice size $L$ in the high temperature phase ($\beta>\beta_c$).
\end{itemize}
As previously discussed these are the properties of $\bar\rho$ which allow to interpret confinement as 
dual superconductivity, as they guarantee that $\langle \bar\mu\rangle\neq 0$ in the low temperature phase and 
$\langle \bar\mu\rangle=0$ in the high temperature one.

\begin{figure}[h]
\scalebox{0.35}{\includegraphics*{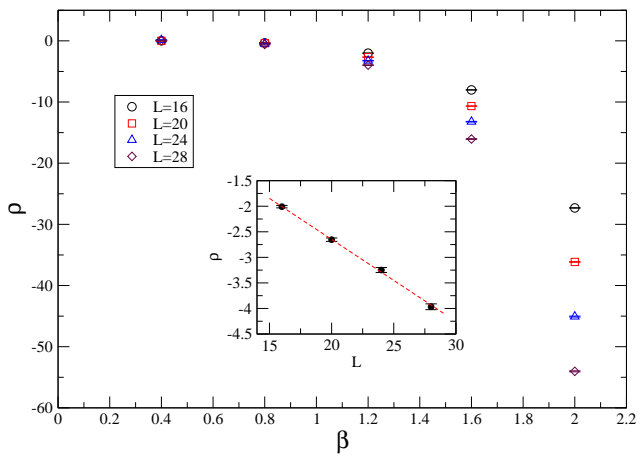}}
\caption{$\rho$ for the $SU(2)$ lattice gauge theory at low $\beta$ calculated for the Wu-Yang monopole of charge 4. 
The inset shows the linear divergence in the lattice size for $\beta=1.2$.}
\label{rho_4xL_low_beta}
\end{figure}

\begin{figure}[h]
\scalebox{0.35}{\includegraphics*{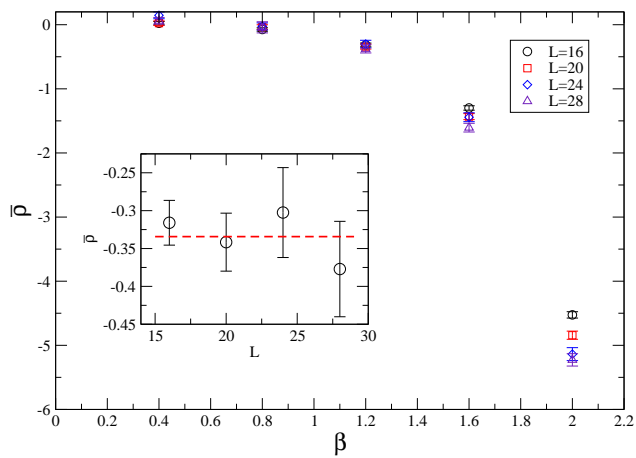}}
\caption{$\bar \rho$ for the $SU(2)$ lattice gauge theory at low $\beta$ calculated for the Wu-Yang monopole of charge 4. 
The inset shows the dependence on the lattice size for $\beta=1.2$ together with the best fit to a constant.}
\label{rhonew_4xL_low_beta}
\end{figure}

\begin{figure}[h]
\scalebox{0.35}{\includegraphics*{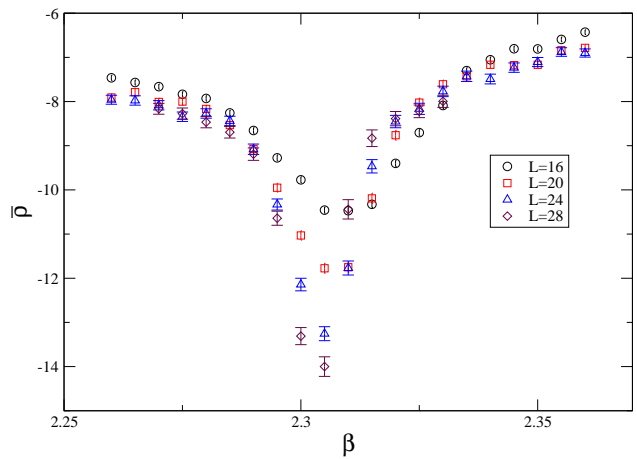}}
\caption{$\bar \rho$ for the $SU(2)$ lattice gauge theory near the deconfinement transition ($N_t=4$) calculated for the 
Wu-Yang monopole of charge 4. When the error bars are not explicitly shown they are smaller than the symbols.}
\label{rhonew_4xL}
\end{figure}

We will first of all study the behaviour of $\rho$ and $\bar{\rho}$ for small values of $\beta$ (\ie $\beta\ll \beta_c$) 
for lattices with temporal extent $N_t=4$ (for which $\beta_c=2.2986(6)$, see \eg \cite{LTW}). The numerical results for 
$\rho$ are shown in Fig.~\eqref{rho_4xL_low_beta} and the linear divergence of $\rho$ with the lattice size (as predicted 
by Eq.~\eqref{rho_div}) is clearly seen. Notice that with lower statistics and smaller lattices the points in the insert of 
Fig.~\eqref{rho_4xL_low_beta} would be undistinguishable within errors as was in Ref.~\cite{dlmp}. 

This infrared divergence is instead absent in the $\bar\rho$ data, which are 
shown in Fig.~\eqref{rhonew_4xL_low_beta} and smoothly approach their thermodynamical values. Clearly this approach 
is slower for the data at larger $\beta$ shown in the figure, which are closer to the deconfinement transition where the
correlation length is larger. Note that, although the absolute values of the error bars for $\rho$ and $\bar\rho$ 
are similar, the relative errors for $\bar\rho$ are much bigger than those for $\rho$ since the value of $\bar\rho$ 
is about an order of magnitude smaller than $\rho$.

The data showing the behaviour of $\bar\rho$ in the neighbourhood of the $N_t=4$ deconfinement transition are displayed in 
Fig.~\eqref{rhonew_4xL} and the development of the negative peak at the transition is clearly visible. The 
deconfinement transition for $3+1$ dimensional $SU(2)$ lattice gauge theory is known to belong to the universality 
class of the 3d Ising model and from a simple scaling ansatz for $\bar\mu$ (see \eg \cite{dp}) the finite size scaling 
relation for the singular part of $\bar\rho$ follows
\begin{equation}\label{rho_scaling}
\bar\rho_{\mathrm{sing}}(\beta)=L^{1/\nu}f\big(L^{1/\nu}(\beta-\beta_c)\big)
\end{equation}
where $f$ is a scaling function. From this scaling form it follows that by increasing the lattice size the peak height 
should increase as $L^{1/\nu}$, while its width should shrink as $L^{-1/\nu}$. This behaviour is qualitatively visible 
in Fig.~\eqref{rhonew_4xL} and will be now verified also quantitatively.

\begin{figure}[h]
\scalebox{0.35}{\includegraphics*{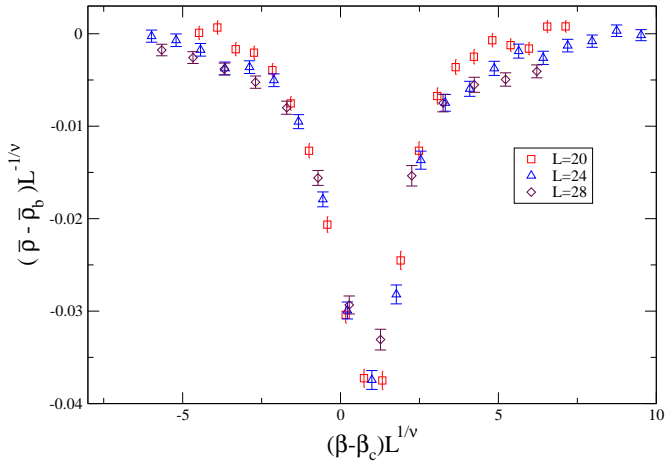}}
\caption{Scaling of the singular part of $\bar\rho$ near the $N_t=4$ deconfinement transition.}
\label{rhonew_4xL_rescaled}
\end{figure}

In order to estimate the analytical background of $\bar\rho$, a fit of the form $a+bL^{1/\nu}$ was performed on the 
peak values of $\bar\rho$ and the constant background term $\bar\rho_{\mathrm{back}}\equiv a$ was used. The quality of 
the scaling in Eq.~\eqref{rho_scaling} is shown in Fig.~\eqref{rhonew_4xL_rescaled}, where the known values 
$\beta_c=2.2986(6)$ and $\nu=0.6301(4)$ where used for the deconfinement coupling and the 3d Ising critical index. 
We explicitly notice that scaling is to be expected only for $\beta\le\beta_c$, since for $\beta>\beta_c$ 
$\bar\rho$ should not be well defined in the thermodynamical limit and it is expected to diverge to $-\infty$ as will be 
directly tested below.

\begin{figure}[h]
\scalebox{0.35}{\includegraphics*{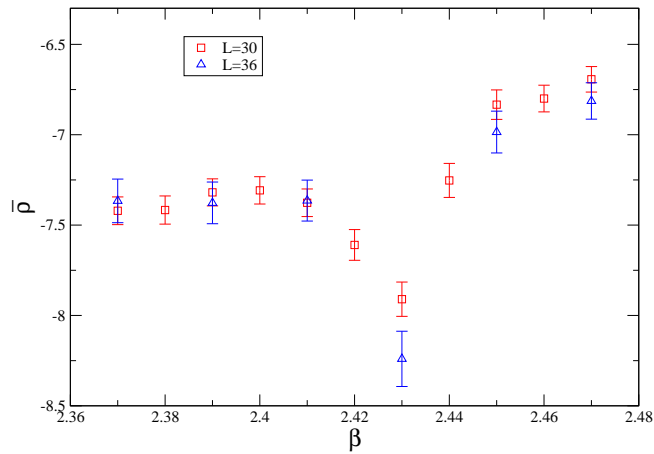}}
\caption{$\bar \rho$ for the $SU(2)$ lattice gauge theory near the deconfinement transition ($N_t=6$) calculated for the 
Wu-Yang monopole of charge 4.}
\label{rhonew_6xL}
\end{figure}

\begin{figure}[ht]
\scalebox{0.35}{\includegraphics*{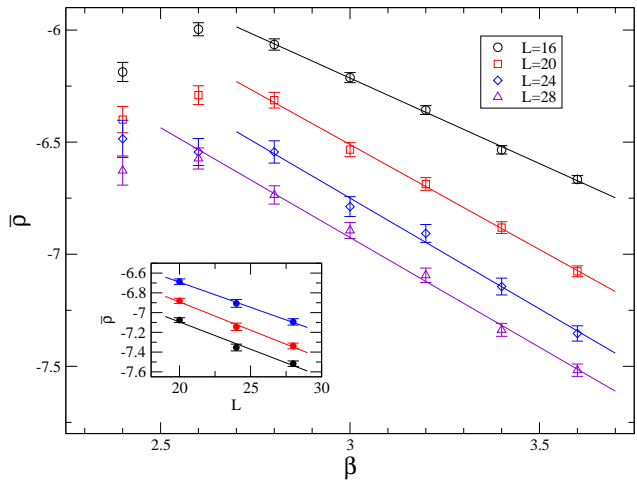}}
\caption{$\bar \rho$ for the $SU(2)$ lattice gauge theory for $\beta>\beta_c$ ($N_t=4$) calculated for the Wu-Yang 
monopole of charge 4. In the inset the scaling with the lattice size at $\beta=3.2, 3.4, 3.6$ is shown together with 
linear fits.}
\label{rhonew_4xL_high_beta}
\end{figure}

To verify that the position of the peak scales correctly with the lattice spacing we simulated also lattices with 
different temporal extent. In Fig.~\eqref{rhonew_6xL} data are shown corresponding to a lattice with $N_t=6$ and 
again the negative peak develops in correspondence of the deconfinement transition, which now takes place at 
$\beta_c=2.4271(17)$ (see \eg \cite{LTW}).

The last point that remains to be checked is the behaviour of $\bar\rho$ at  temperatures higher than the deconfinement temperature.
 In order to have $\langle\bar\mu\rangle=0$ for $\beta>\beta_c$, $\bar\rho$ must diverge to $-\infty$ in the 
thermodynamical limit for every $\beta>\beta_c$. This is indeed what happens, as shown in Fig.~\eqref{rhonew_4xL_high_beta},
where two different scaling are seen: at fixed lattice size $\bar\rho$ is asymptotically linear in $\beta$ while at fixed 
$\beta$ value $\bar\rho$ is asymptotically linear in the lattice size $L$.

We have thus verified numerically that the improved monopole operator $\bar\mu$ is a well defined operator to check dual 
superconductivity in lattice simulations. We have thus shown that the deconfinement transition in $SU(2)$ lattice gauge 
theory can be  interpreted as a monopole condensation transition.

\section{Subtraction at zero temperature.}\label{sub_sec}

\begin{figure}[h]
\scalebox{0.35}{\includegraphics*{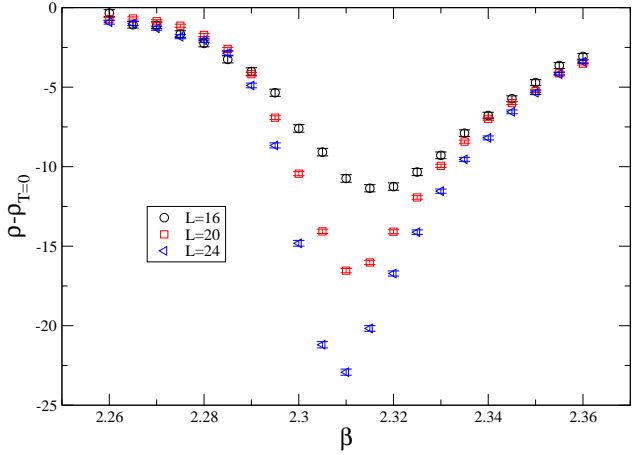}}
\caption{The peak in $\rho-\rho_{T=0}$ near $\beta_c$.}
\label{rhosub_4xL}
\end{figure}

\begin{figure}[h]
\scalebox{0.35}{\includegraphics*{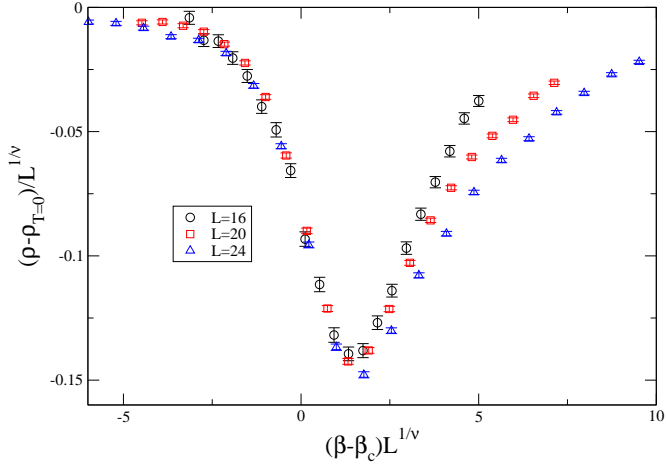}}
\caption{Scaling of $\rho-\rho_{T=0}$ at deconfinement.}
\label{rhosub_4xL_rescaled}
\end{figure}

For the sake of completeness we also report some numerical results of an attempt we made prior to the approach described in
this paper on the way to understand the diseases of our disorder parameter \cite{lat}.
The basic idea was that the sort of infrared divergence appearing in $\rho$ would look as a short distance effect in the 
dual description and produce a multiplicative renormalization of $\langle \mu \rangle$. In the ratio $\frac{\mu_T}{\mu_{T=0}}$ 
the constant would cancel and the deconfining transition become visible. The analogue of $\bar{\rho}$ becomes now $\rho - \rho_{T=0}$. 
The results are shown in Figs.~\eqref{rhosub_4xL} and \eqref{rhosub_4xL_rescaled}. We report these results because they prove 
to be meaningful. 

\begin{figure}[h]
\scalebox{0.35}{\includegraphics*{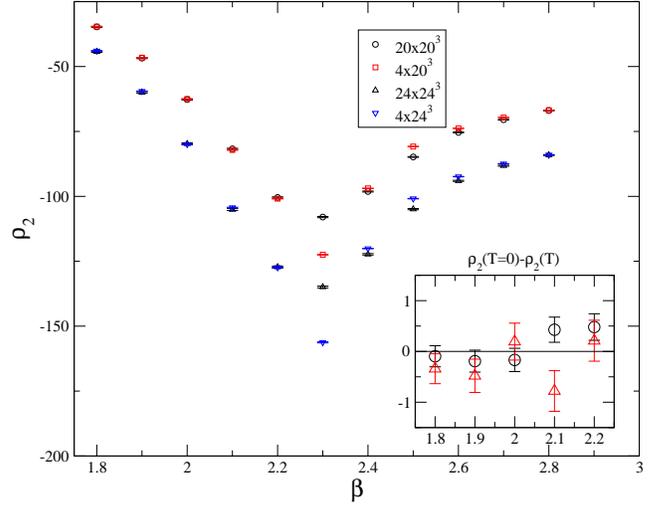}}
\caption{$\rho_2$ is T-independent at low $\beta$. As a consequence $\langle \mu\rangle/\langle \mu_{T=0}
\rangle=\langle \bar{\mu}\rangle/\langle \bar{\mu}_{T=0}\rangle$.}
\label{rho2_zeroT_finiteT}
\end{figure}

It is not easy to simulate the system at $T=0$ and large $\beta$, since the physical length grows exponentially with $\beta$ 
and the lattice size should be increased accordingly. In the practically accessible range of values of $\beta$ the results 
are satisfactory: the phase transition is clearly visible and the appropriate scaling is observed. In terms of the new 
disorder parameter this means that $\rho_2$ is rather smooth in $\beta$ and temperature independent below and around 
the deconfinement peak. This can indeed be directly seen in Fig.~\eqref{rho2_zeroT_finiteT} where the term $\rho_2$ of 
Eq.~\eqref{ro2}, which determines the normalization Eq.~\eqref{mus}, is shown to be $N_t$ or temperature independent 
in a range of $\beta$'s.

\begin{figure}[h]
\scalebox{0.35}{\includegraphics*{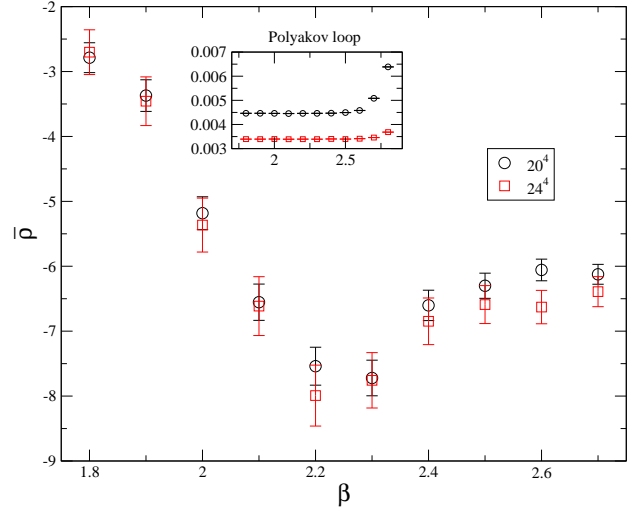}}
\caption{$\bar{\rho}$ in finite and volume independent at $T=0$: this implies that $\langle\bar\mu\rangle\neq 0$
and there is confinement.}
\label{rhonew_zeroT}
\end{figure}

The behavior of $\bar\rho$ at zero temperature is also shown in Fig.~\eqref{rhonew_zeroT} to be volume independent and 
finite, meaning that there is confinement.


\section{Conclusions and outlook}\label{concl_sec}
Understanding the mechanism of confinement of color in $QCD$ is a fundamental problem in particle physics. Dual 
superconductivity of the confining vacuum is an appealing candidate.

A disorder parameter had been introduced to detect condensation of magnetic charges, in order to demonstrate by numerical 
simulations that deconfinement is a transition from a dual superconductor to a normal state. The starting point was the 
$U(1)$ lattice gauge theory which has a deconfining transition well understood in terms of dual superconductivity. The 
generalization of the $U(1)$ disorder parameter to the non-abelian case proved to be affected by a pathology: in the 
thermodynamic limit the parameter as was constructed tends to zero both in the deconfined and in the confined phase, 
and is thus unable to detect superconductivity.

In this paper we trace the origin of this disease and we cure it. In synthesis the original disorder parameter was not 
a properly normalized probability amplitude, and had limit zero at increasing volumes because the norm of the state obtained 
from the vacuum by addition of a monopole tends to zero. The direction of the state in the Hilbert space has instead a well 
defined limit, and with it the probability amplitude in the confined phase.  This has been shown both by strong coupling
expansion and by numerical simulations.
 
The results support dual superconductivity as mechanism of color confinement.
 
The disorder parameter is by construction gauge invariant if the classical field of the monopole transforms covariantly. 
In other words it is not invariant under gauge transformations of the classical external field. This is not a difficulty in 
principle for the demonstration of monopole condensation, but is unpleasant.
 
We are trying to use alternative definitions, as could be the technique of the Schroedinger  functional, which was already used 
in Ref.~\cite{ccos} which could be more satisfactory.

\section{aknowledgments}
One of us (A.~D.~G.) thanks G.~Paffuti for useful discussions. The numerical simulations have been performed on GRID 
resources provided by INFN and in particular on the CSN4 Cluster located in Pisa.

\appendix
\section{Strong coupling expansion}\label{strong_sec}

In this section we sketch the strong coupling expansion of $\rho$ and $\bar{\rho}$ of Sec. \ref{impro_sec}. 
The strong coupling expansion is a power series expansion in $\beta$ (see \eg \cite{Cbook, MMbook}).
More refined methods, such as the character expansion (see \cite{MMbook}), cannot be used here because of the 
monopole insertion in $S+\Delta S$.

For the $U(1)$ case we want to compute the expansion of 
\begin{equation}
\rho=\langle S\rangle_S-\langle S+\Delta S\rangle_{S+\Delta S} \label{rho_app}
\end{equation}
where $S$ and $S+\Delta S$ are defined in Eqs.\eqref{S_ab}-\eqref{pp}. The fundamental formulae to be used are
\begin{equation}
\begin{aligned} \label{ab_rules}
& \int_0^{2\pi}\frac{\mathrm{d}x}{2\pi}\cos(a+x)=0 \\
& \int_0^{2\pi}\frac{\mathrm{d}x}{2\pi}\cos(a+x)\cos(x-b)=\frac{1}{2}\cos(a+b)
\end{aligned}
\end{equation}
It has to be noted that in the expression \eqref{rho_app} most of the terms cacels between 
$\langle S\rangle_S$ and $\langle S+\Delta S\rangle_{S+\Delta S}$. 
For example the lowest non-trivial contribution to both $\langle S\rangle_S$ and 
$\langle S+\Delta S\rangle_{S+\Delta S}$ are $O(\beta)$ and can be graphically represented by two superimposed plaquettes 
(if the two plaquettes are not superimposed the result is zero because of the first of Eqs.~\eqref{ab_rules}). 
However these two contributions to $\langle S\rangle_S$ and $\langle S+\Delta S\rangle_{S+\Delta S}$ are equal and cancel 
in Eq.~\eqref{rho_app}.

\begin{figure}[h]
\centering
\scalebox{0.2}{\includegraphics*{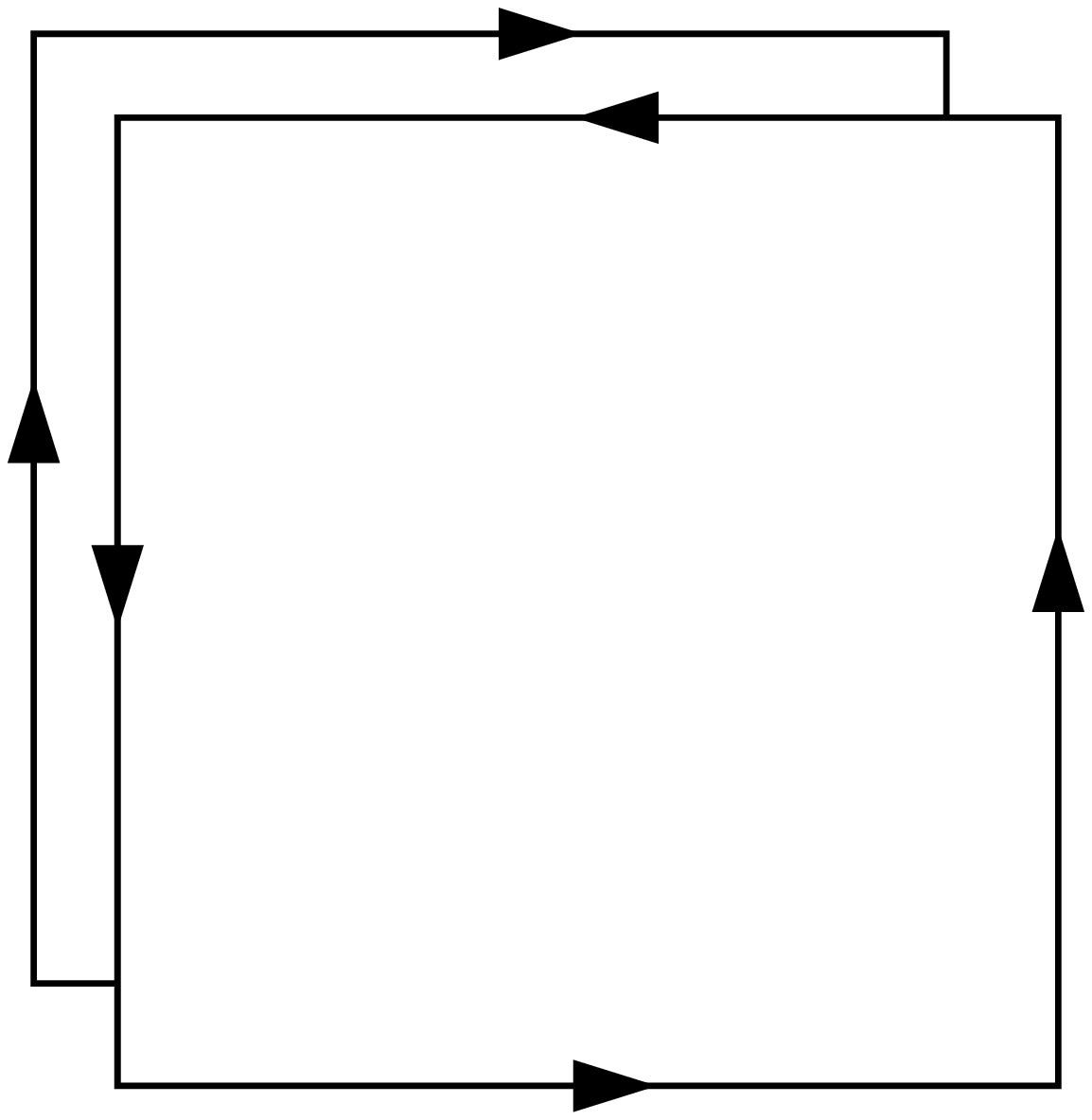}}
\hspace{1cm}
\scalebox{0.2}{\includegraphics*{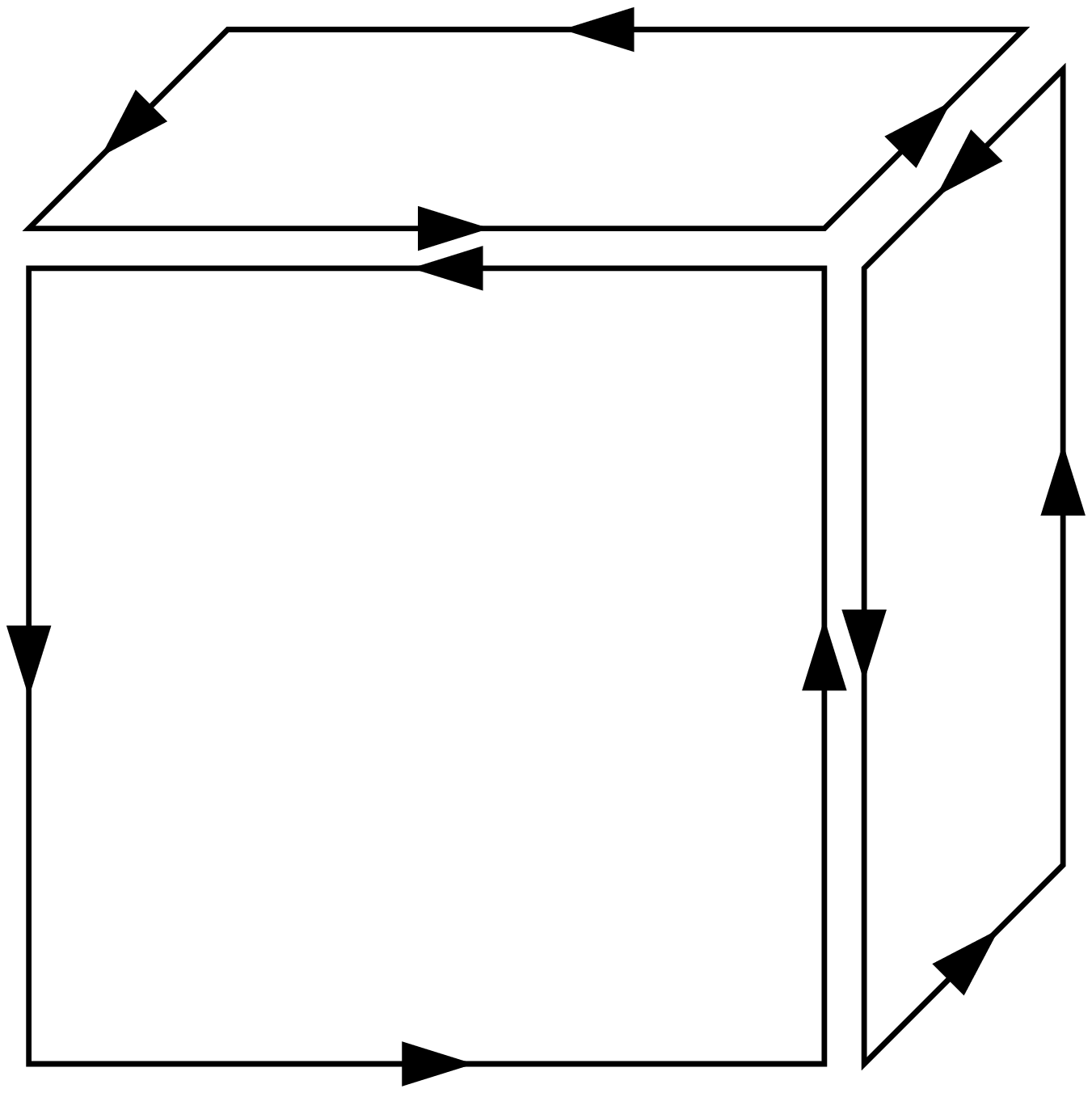}}
\caption{(\emph{left}) Contribution of $O(\beta)$ to $\langle S\rangle_S$ but not to $\rho$.
(\emph{right}) Contribution of $O(\beta^5)$ to both $\langle S\rangle_S$ and $\rho$.}
\label{strong_fig}
\end{figure}

The first non-trivial contribution to $\rho$ turns out to be $O(\beta^5)$ and is graphically represented by the six 
plaquettes that enclose a three-dimensional cube. If the cube has no edges along the temporal direction and/or if it 
is not positioned at the time at which the monopole is created, it simplifies in the difference in Eq.~\eqref{rho_app}. 

Let us denote by $T_{ij}(\vec n)$ the contribution to the mean value $\langle S+\Delta S\rangle_{S+\Delta S}$
associated to a cube sitting at $(0,\vec{n})$ with edges along the directions $\hat{0}, \hat{\imath}, 
\hat{\jmath}$ ($0$ is the temporal direction). By applying 12 times (the number of the edges of the cube) 
the second formula of Eqs.~\eqref{ab_rules}
we get
\begin{equation}
T_{ij}=-\frac{1}{2^{12}}\cos\big(F_{ij}^{0}(\vec{n})\big)
\end{equation}
from which Eq.~\eqref{sc} follows.

In the non-abelian case the integration rules Eqs.~\eqref{ab_rules} become 
\begin{equation}\label{nonab_rules}
\begin{aligned}
& \int\mathrm{d}U\, U_{ij}=0 \\
& \int\mathrm{d}U\, U_{ij}(U^{-1})_{kl}=\frac{1}{\Tr(I)}\delta_{jk}\delta_{il}
\end{aligned}
\end{equation}
and the computations goes along the same line. Also in the non-abelian case the first non-trivial 
contribution to $\rho$ is $O(\beta^5)$ and correspond to the cube graph of in Fig.~\eqref{strong_fig}.

By using the second relation of Eqs.~\eqref{nonab_rules} on the twelve edges of the cube we obtain the expressions 
in Eq.~\eqref{osc} and Eq.~\eqref{osc2}.

\section{Details of the gauge transformation of Section \ref{mono_sec}.}\label{gauge_change_sec}

In our notation $\sigma_r, \sigma_{\theta}, \sigma_{\phi} $ are the three generators along the axes of the polar 
coordinates. In terms of the generators along the cartesian axes, $\sigma_1, \sigma_2, \sigma_3 $,they are given by 
\begin{equation}
\begin{aligned}
\sigma_r =  \sin \theta (\sigma_1 \cos \phi + \sigma_2  \sin\phi) + \sigma_3 \cos\theta\\
\sigma_{\theta} = \cos \theta (\sigma_1 \cos \phi + \sigma_2  \sin\phi) - \sigma_3 \sin\theta\\
\sigma_{\phi} =  - \sigma_1 \sin\phi + \sigma_2 \cos\phi
\end{aligned}
\end{equation}
By use of Eq.~\eqref{u} the covariant terms in Eq.~\eqref{A} are easily computed giving
\begin{equation}
U(\theta,\phi)^{\dagger}  A_{r} U(\theta,\phi) =0
\end{equation}
\begin{equation}
U(\theta,\phi)^{\dagger}  A_{\theta} U(\theta,\phi) =
\frac{1}{2gr}\Lambda^{\dagger} \sigma_{\phi} \Lambda
\end{equation}
\begin{equation}
\begin{split}
& U(\theta,\phi)^{\dagger} A_{\phi} U(\theta,\phi)= \\
& \hspace{2cm}=-\frac{1}{2gr} \Lambda^{\dagger} (\sigma_{\bar \Theta} \cos\theta + 
\sigma_r \sin{\bar \Theta}) \Lambda
\end{split}
\end{equation}
The affine terms $-\frac{i}{g}  U(\theta,\phi)^{\dagger} \vec \nabla U(\theta,\phi)$ are
\begin{equation}
-\frac{i}{g} U(\theta,\phi)^{\dagger} \partial_{r} U(\theta,\phi) =0
\end{equation}
\begin{equation}
\begin{split}
& -\frac{i}{g} U(\theta,\phi)^{\dagger}\frac{1}{r} \partial_{\theta} U(\theta,\phi) =\\ 
&\hspace{1cm}=-\frac{\bar \Theta'}{2gr} \Lambda^{\dagger} \sigma_{\phi} \Lambda -\frac{i}{gr} 
\Lambda^{\dagger} \partial_{\theta}\Lambda
\end{split}
\end{equation}
\begin{equation}
\begin{split}
& -\frac{i}{g} U(\theta,\phi)^{\dagger}\frac{1}{r \sin {\theta}} \partial_{\phi} U(\theta,\phi)= \\
&\hspace{1cm} =\frac{1}{2gr \sin\theta}\Lambda^{\dagger} \[\sin{\bar \Theta}(\sigma_{\theta} \cos\theta + 
\sigma_{r} \sin\theta) +\\
&\hspace{4cm} +(1- \cos{\bar \Theta}) \sigma_3\] \Lambda
\end{split}
\end{equation}
Using the fact that $\bar \Theta ' \equiv \frac{d\bar \Theta}{d\theta} = 1 - \pi \delta(\theta - \theta_{WY}) $, 
the result for $\vec{\tilde A}$ is that of  Eq.'s~\eqref{Ar}\eqref{Ateta}\eqref{Afi}. 

We now compute the magnetic field by Eq.~\ref{mf}. The three terms give separately
\begin{equation}
\begin{split}
T_1 &\equiv  \frac{1}{r \sin\theta}\partial_{\theta}( \sin\theta  {\tilde A}_{\phi}) =  \\
&=\frac{1}{2r^2 \sin\theta}\partial_{\theta}[(1- \cos{\bar \Theta)}\Lambda^{\dagger} \sigma_{3}\Lambda]
\end{split}
\end{equation}
The term $\cos{\bar \Theta}\Lambda^{\dagger} \sigma_{3}\Lambda = \cos\theta \sigma_{3}$ is continuous through $\theta_{WY}$ and gives the 
coulombic field of the monopole. Finally
\begin{equation}
T_1 = \frac{\sigma_3}{2} \frac{1}{gr^2} +\frac{ \pi \delta(\theta - \theta_{WY})}{2gr^2\sin\theta}\Lambda^{\dagger}
\frac{-i}{2}[ \sigma_{\perp},\sigma_{3}]\Lambda
\end{equation}
Since $\partial_{\phi} \sigma_{\phi} = -(\sigma_{r}\sin\theta + \sigma_{\theta} \cos\theta)$, the second term in 
Eq.~\eqref{mf} gives
\begin{equation}
\begin{split}
T_2 &\equiv -\frac{1}{r\sin\theta} \partial_{\phi} \tilde A_{\theta} = \\ 
&=\frac{ \pi \delta(\theta - \theta_{WY})}{2gr^2
\sin\theta} \Lambda^{\dagger}(\sigma_{r}\sin\theta + \sigma_{\theta} \cos\theta)\Lambda
\end{split}
\end{equation}
Finally the bilinear term gives
\begin{equation}
\begin{split}
T_3 &\equiv ig[\tilde A_{\theta}, \tilde A_{\phi}] = \\
&=\frac{ \pi \delta(\theta - \theta_{WY})}{2gr^2\sin\theta}(1- \cos{\bar \Theta})\times \\
&\hspace{2cm}\times \Lambda^{\dagger}\frac{i}{2}[(\sigma_{\phi}+ \sigma_{\perp}),\sigma_3]\Lambda
\end{split}
\end{equation}
Since $\cos \bar{\Theta}\, \delta(\theta - \theta_{WY}) =0$ and 
\begin{equation}
[\sigma_{3},\sigma_{\phi}] = - 2i ( \sigma_{\theta} \cos\theta + \sigma_{r} \sin\theta)
\end{equation}
we get
\begin{equation}
\begin{split}
T_3 &= \frac{ \pi \delta(\theta - \theta_{WY})}{2gr^2\sin\theta} \Lambda^{\dagger} \big\{ -( \sigma_{\theta} \cos\theta +\sigma_{r} \sin\theta)+\\
&\hspace{4cm} +\frac{i}{2}[ \sigma_{\perp},\sigma_3]\big\}\Lambda
\end{split}
\end{equation}
All the singularities cancel in the sum and only the coulombic term is left.

Notice that, since $\bar \sigma \equiv \frac{-i}{2}[\sigma_{\perp}, \sigma_3]$ is orthogonal to $\sigma_{\perp}$ 
\begin{equation}
\begin{split}
&\Lambda^{\dagger}\frac{-i}{2}[ \sigma_{\perp},\sigma_{3}]\Lambda\delta(\theta - \theta_{WY})= \\
&\quad=\bar \sigma \cos (\bar \Theta -\theta)\delta(\theta - \theta_{WY}) =0 \label{ub}
\end{split}
\end{equation}
The discontinuity in $T_1$ vanishes, and so does the second term of $T_3$ which should cancel it.


\begin{thebibliography}{99}
 \bibitem{'tHP} G.~'t~Hooft, in \emph{High Energy Physics}, EPS International Conference, Palermo 1975, A. Zichichi ed.
 (Editrice Compositori, Bologna, 1976).
 \bibitem{m} S.~Mandelstam, Phys. Rep. {\bf 23}, 245 (1976).
 \bibitem{'tH2} G.~'t~Hooft, Nucl. Phys. B {\bf 190}, 455 (1981).
 \bibitem{sco}  T.~Sekido, K.~Ishiguro,Y.~Koma, Y.~Mori, T.~Suzuki, Phys. Rev. D {\bf 76}, 031501 (2007) [arXiv:hep-lat/0703002].
 \bibitem{suz} T.~Suzuki, I.~Yotsuyanagi, Phys. Rev. D {\bf 42}, 4257 (1990).
 \bibitem{pol} M.~I.~Polikarpov, Nucl. Phys. B (Proc. Suppl.) {\bf 53}, 134 (1997) [arXiv:hep-lat/9609020].
 \bibitem{ts} T. Suzuki et al. Nucl. Phys. B (Proc. Suppl.) {\bf 53}, 531 (1997) [arXiv:hep-lat/9607054].
 \bibitem{bdlp} C.~Bonati, A.~Di~Giacomo, L.~Lepori, F.~Pucci, Phys. Rev. D {\bf 81}, 085022 (2010) [arXiv:1002.3874].
 \bibitem{bdd} C.~Bonati, A.~Di~Giacomo, M.~D'Elia, Phys. Rev. D {\bf 82}, 094509 (2010) [arXiv:1009.2425].
 \bibitem{dgt} T.~A.~DeGrand, D.~Toussaint, Phys. Rev. D {\bf 22}, 2478 (1980).
 \bibitem{digz} A.~Di~Giacomo, Acta Phys. Polon. B {\bf 25}, 215 (1994). 
 \bibitem{dp} A.~Di~Giacomo, G.~Paffuti, Phys. Rev. D {\bf 56}, 6816 (1997) [arXiv:hep-lat/9707003].
 \bibitem{fm} J.~Fr\"{o}hlich, P.~Marchetti, Comm. Math. Phys. {\bf 112}, 343 (1987).
 \bibitem{cp} V.~Cirigliano, G.~Paffuti, Commun. Math. Phys. {\bf 200}, 381 (1999).
 \bibitem{dirac} P. A. M.~Dirac, Canad. Journ. of Phys. {\bf 33}, 650 (1955).
 \bibitem{cd} J.~M.~Carmona, A.~Di Giacomo, Phys. Lett. B {\bf 485}, 126 (2000) [arXiv:hep-lat/0005014].
 \bibitem{ddpt} G.~Di Cecio, A.~Di Giacomo, G.~Paffuti, M.~Trigiante, Nucl. Phys. B {\bf 489}, 739 (1997) [arXiv:hep-lat/9608014].
 \bibitem{dmp} A.~Di Giacomo, D.~Martelli, G.~Paffuti, Phys. Rev. D {\bf 60}, 094511 (1999) [arXiv:hep-lat/9905007].
 \bibitem{dlmp} A.~Di~Giacomo, B.~Lucini, L.~Montesi, G.~Paffuti. Phys. Rev. D {\bf 61}, 034503 (2000) [arXiv:hep-lat/9906024].
 \bibitem{ccos} P.~Cea, L.~Cosmai, Phys. Rev. D {\bf 62}, 094510 (2000) [arXiv:hep-lat/0006007].
 \bibitem{dlmp1} A.~Di~Giacomo, B.~Lucini, L.~Montesi, G.~Paffuti. Phys. Rev. D {\bf 61}, 034504 (2000) [arXiv:hep-lat/9906025].
 \bibitem{3} J.~M.~Carmona, M.~D'Elia, A.~Di~Giacomo, B.~Lucini, G.~Paffuti, Phys. Rev. D {\bf 64}, 114507 (2001) [arXiv:hep-lat/0103005].
 \bibitem{4} J.~M.~Carmona, M.~D'Elia, L.~Del~Debbio, A.~Di~Giacomo, B.~Lucini, G.~Paffuti,
Phys. Rev. D {\bf 66}, 011503 (2002) [arXiv:hep-lat/0205025].
 \bibitem{vpc}A.~I.~Veselov, M.~I.~Polikarpov, M.~N.~Chernodub, JETP Lett. {\bf 63}, 411 (1996), Pisma Zh. Eksp. 
Teor. Fiz. {\bf 63}, 392 (1996). 
 \bibitem{cddlp} G.~Cossu, M.~D'Elia, A.~Di~ Giacomo, B.~Lucini, C.~Pica, PoS LAT2006:063(2006) [arXiv:hep-lat/0609061] and unpublished.
 \bibitem{pw} K.~Holland, P.~Minkowski, M.~Pepe, U.~J.~Wiese, Nucl. Phys. B {\bf 668}, 207 (2003) [arXiv:hep-lat/0302023].
 \bibitem{pw1} M.~Pepe, Nucl. Phys. B (Proc. Suppl.) {\bf 153}, 207 (2006) [arXiv:hep-lat/0510013]. 
 \bibitem{go} J.~Greensite, K.~Langfeld, S.~Olejnik, H.~Reinhardt, T.~Tok, Phys. Rev. D {\bf 75}, 034501 (2007) [arXiv:hep-lat/0609050].
 \bibitem{dlp} A.~Di~Giacomo, L.~Lepori, F.~Pucci, JHEP {\bf 0810}, 096 (2008) [arXiv:0810.4226].
 \bibitem{gl} J.~Greensite, B.~Lucini, Phys. Rev. D {\bf 78}, 085004 (2008) [arXiv:0806.2117].
 \bibitem{fmb} J.~Fr\"{o}hlich, P.~A.~Marchetti, Phys. Rev. D {\bf 64}, 014505 (2001) [arXiv:hep-th/0011246]. 
 \bibitem{'tHm} G.~'t~Hooft, Nucl. Phys. B {\bf 79}, 276 (1974).
 \bibitem{Pol} A.~M.~Polyakov, JETP Lett.  {\bf 20}, 194 (1974).
 \bibitem{lat} C.~Bonati, G.~Cossu, M.~D'Elia, A.~Di Giacomo, PoS LATTICE2010:271(2010) [arXiv:1010.5428].
 \bibitem{w} K.~Wilson, Phys. Rev. D {\bf 10}, 2445 (1974).
 \bibitem{shnir} Ya.~Shnir, ``Magnetic Monopoles'', Springer (2005).
 \bibitem{creutz80} M.~Creutz, Phys. Rev. D {\bf 21}, 2308 (1980).
 \bibitem{kp} A.~D.~Kennedy, B.~J.~Pendleton, Phys. Lett. B {\bf 156}, 393 (1985).
 \bibitem{creutz87} M.~Creutz, Phys. Rev. D {\bf 36}, 515 (1987).
 \bibitem{LTW} B.~Lucini, M.~Teper, U.~Wenger, JHEP {\bf 0401}, 061 (2004) [arXiv:hep-lat/0307017].
 \bibitem{Cbook} C.~Creutz, ``Quarks, gluons and lattices'', Cambridge University Press (1983).
 \bibitem{MMbook} I.~Montvay, G.~M\"{u}nster, ``Quantum fields on a lattice'', Cambridge University Press (1994).
 \end{thebibliography}
\end{document}